\documentclass[conference]{IEEEtran}
\IEEEoverridecommandlockouts
\usepackage{soul}
\usepackage{cite}
\usepackage{amsmath,amssymb,amsfonts}
\usepackage{algorithmic}
\usepackage{graphicx}
\usepackage{textcomp}
\usepackage[dvipsnames]{xcolor}
\usepackage{array}
\usepackage{todonotes}

\usepackage[utf8]{inputenc}
\usepackage{multirow}

\usepackage{tabularx}
\usepackage{multirow}
\usepackage{pgfplots}
\usepackage{lipsum}
\usepackage{cancel}
\usepackage[top=2.5cm,bottom=2.5cm,left=3cm,right=2cm]{geometry}
\usepackage{marginnote}

\usepackage{bm}

\usepackage{ulem}
\usepackage{amsmath,amsfonts,amssymb}

\usepackage[style=plain]{floatrow}
\usepackage{graphicx}
\usepackage{subfig}

\pgfplotsset{} 
\def\BibTeX{{\rm B\kern-.05em{\sc i\kern-.025em b}\kern-.08em
    T\kern-.1667em\lower.7ex\hbox{E}\kern-.125emX}}

\pgfplotsset{} 

\makeatletter
\newcommand{\linebreakand}{%
  \end{@IEEEauthorhalign}
  \hfill\mbox{}\par
  \mbox{}\hfill\begin{@IEEEauthorhalign}
}
\makeatother

\begin{document}



\title{Comparing Autoencoder to Geometrical Features for Vascular Bifurcations Identification}


\author{\IEEEauthorblockN{Ibtissam Essadik}
\IEEEauthorblockA{\textit{SETIME Laboratory} \\
\textit{Faculty of Science, Ibn Tofail University}\\
Kenitra, Morocco \\
ibtissam.essadik@uit.ac.ma}
\and
\IEEEauthorblockN{Anass Nouri}
\IEEEauthorblockA{\textit{SETIME Laboratory} \\
\textit{Faculty of Science, ENSC, Ibn Tofail University}\\
Kenitra, Morocco \\
anass.nouri@uit.ac.ma}
\linebreakand
\IEEEauthorblockN{Raja Touahni}
\IEEEauthorblockA{\textit{SETIME Laboratory} \\
\textit{Faculty of Science, Ibn Tofail University}\\
Kenitra, Morocco\\
raja.touahni@uit.ac.ma}
\and
\IEEEauthorblockN{Florent Autrusseau}
\IEEEauthorblockA{\textit{Institut du Thorax and LTeN Labs} \\
\textit{University of Nantes}\\
Nantes, France \\
Florent.Autrusseau@univ-nantes.fr}}

\maketitle

\begin{abstract}

The cerebrovascular tree is a complex anatomical structure that plays a crucial role in the brain irrigation. A precise identification of the bifurcations in the vascular network is essential for understanding various cerebral pathologies. Traditional methods often require manual intervention and are sensitive to variations in data quality. In recent years, deep learning techniques, and particularly autoencoders, have shown promising performances for feature extraction and pattern recognition in a variety of domains. In this paper, we propose two novel approaches for vascular bifurcation identification based respectiveley on Autoencoder and geometrical features. The performance and effectiveness of each method in terms of classification of vascular bifurcations using medical imaging data is presented. The evaluation was performed on a sample database composed of 91 TOF-MRA, using various evaluation measures, including accuracy, F1 score and confusion matrix.

\end{abstract}


\section{Introduction}
Recognizing the bifurcations composing the Circle of Willis (CoW) may play an important role in the diagnostic of various cerebrovascular diseases such as intracranial aneurysms. The CoW is a complex vascular structure formed by over 30 different arteries, merging into more than 15 bifurcations. An accurate identification of the various portions constituting the CoW is crucial to understand the blood flow along different parts of the brain and to detect any abnormalities.
Anatomical labeling can be considered as the mapping of an unlabeled case to an atlas, represented by a knowledge base of the population mean with geometrical and structural variability.  

Several initiatives have been undertaken for the development of brain atlases, aiming to identify distinct bifurcations in the cerebrovascular system. Bogunovic \textit{et al.}~\cite{b1} proposed a method based on \textit{a posteriori} maximum likelihood estimation to automatically assign labels to the various branches and bifurcations of the CoW. Their solution combined both anatomical knowledge and bifurcation characteristics.
In another work conducted by Bilgel \textit{et al.}~\cite{b2}, the authors proposed a method for labeling the anterior part of the cerebral vasculature using belief propagation based on a Bayesian model. Ghanavati \textit{et al.}~\cite{b3} labeled the whole mouse circulatory system. The labeling problem was expressed as a Markov random field, where the best solution was obtained through simulated stochastic relaxation annealing. In \cite{b4}, Robben \textit{et al.} employed graphical matching techniques to locate and identify human brain arteries.
Artificial intelligence algorithms, particularly machine learning and deep learning, are widely used in the field of medical imaging analysis. Some studies use machine learning algorithms to detect and classify CoW bifurcations and/or arteries. Murat \textit{et al.}~\cite{b5}  used a random forest and a Bayesian network to identify brain arteries. Wang \textit{et al.}~\cite{b6} employed supervised machine learning methods to automatically classify brain arteries. Their approach first extracts geometric bifurcations from the vessel centerlines, from which a collection of geometric features is derived. The probability distribution of each bifurcation is then trained using an XGBoost classifier. Finally, a hidden Markov model with a constrained transition strategy is constructed to find the most likely labeling configuration of the entire structure while ensuring topological consistency. In \cite{b7}, Ota \textit{et al.} have used the AdaBoost multi-class methodology to issue the anatomical label of the CoW branches.
In one of our previous works \cite{b8}, we have investigated the labeling of the cerebral bifurcations using geometrical arteries features in order to design an algorithm being able to automatically identify the main bifurcations constituting the CoW. The proposed solution was based on dimensionality reduction and machine learning algorithms. Our method was evaluated using two types of 3D vascular images: A synthetic tree and mouse brain vasculatures. 

In this study, we introduce a new automatic technique to identify various bifurcations of interest within human brain vasculature images. 
We feed our classifiers with a wide variety of cerebral bifurcations (some do form the CoW, some do not), and hence, our approach does not need any prior bifurcation extraction.
This research is a further development of our previous work \cite{b9} dealing with the automatic recognition of primary bifurcations constituting the CoW in human MRA data. To the best of our knowledge, this is the first attempt to design and test a combination of 3D autoencoder for features extraction/reduction and machine learning for classification of bifurcations using MRA images. Unlike previous works \cite{b1,b4,b6}, our method allows to classify 13 bifurcations of interest gathered from the entire vascular tree. 
Indeed, so far most works from the literature were able to classify the bifurcations composing the CoW, but, they all came with the same drawback: The need to have a prior selection of the bifurcations of interest. These methods are not capable of recognizing the bifurcations of interest out of the entire image. Our method can accurately recognize the main CoW bifurcations without any prior processing.

Furthermore, in this work, not only do we consider the geographic location and the full geometrical characterization of the bifurcations, but we also take advantage of a 3D autoencoder for image feature extraction by compressing the input bifurcation into a lower dimensional representation. The compressed representation would be used as the feature vector in the classification task. 

\begin{figure} [ht]
   \begin{center}
   \includegraphics[height=5.1cm]{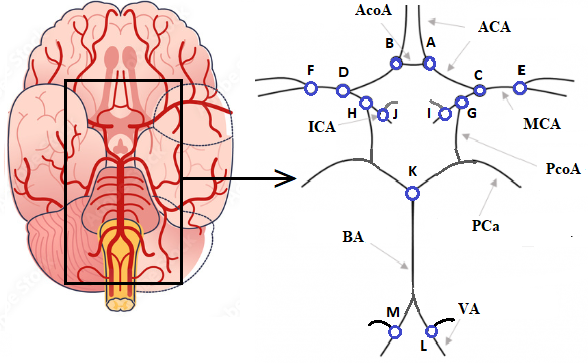}
   \end{center}
   \caption{\label{fig:CoW}Schematic representation of the BoIs constituting the CoW.}
\end{figure}

The CoW is formed by the fusion of several major arteries, including the internal carotid arteries (ICaA), basilar artery (BA) and vertebral artery (VA).  Other CoW arteries include the anterior cerebral arteries (ACA), middle cerebral arteries (MCA), posterior cerebral arteries (PCa), anterior communicating artery (AcoA) and posterior communicating arteries (PcoA). We use labels A to M for the 13 BoIs (see Fig.~\ref{fig:CoW}).

Moreover, it is crucial to consider all the cerebral bifurcations (\textit{i.e.}, even those being outside the CoW) within the classification process. From now on, such bifurcations will be denoted as bifurcations of no interest (BNs). This class was thus added to allow the classifier to recognize the bifurcations of interest among all the cerebral bifurcations. 

In the next sections, we describe our proposed methods for labeling the cerebral bifurcations and the different necessary preprocessing steps. Then, we evaluate the approaches and demonstrate their advantages over current techniques. Finally, we conclude this work and discuss possible improvements.

\section{Methodology}

In this work, we propose machine learning and deep learning-based methods for the analysis of the human brain and a joint detection and classification of the major cerebrovascular bifurcations where aneurysms frequently occur. Brain images are acquired through Magnetic Resonance Angiography and 
a Time Of Flight modality (MRA-TOF). 

The proposed approach is divided into two main parts: feature extraction and classification. Feature extraction is critical in machine learning classification since it allows the recognition of patterns and discriminative representations in complex data, such as bifurcation segments in our context. Autoencoders have received much attention in the recent years due to their ability to construct accurate representations from unlabeled inputs \cite{b14}. Therefore, this study contributes to the understanding of the impact of autoencoders on feature extraction from 3D bifurcation patches. We highlight the potential of autoencoders to discover latent representations and improve the accuracy of bifurcation classification and identification. 
In the classification phase, different machine learning and deep learning algorithms are trained using labeled datasets of brain images with known bifurcations and aneurysms. The classifiers are evaluated on the basis of their accuracy, F1-score, and other performance metrics.  

In this work, we intend to propose a comparative study between autoencoder features and geometrical features for the task of identifying human brain bifurcations with different supervised classification algorithms. The following subsection presents our proposed methods for these two scenarios.

\subsection{Classification based on geometrical features}

\begin{figure} [ht]
   \begin{center}
   \includegraphics[height=5.2cm]{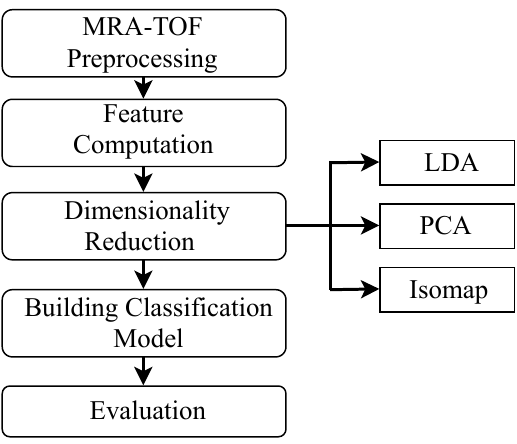}
   \end{center}
   \caption{\label{fig:schema_DR}Overview of the proposed methodology based on features geometrical}
\end{figure}

In this section, we present the proposed methodology for automatic labeling of cerebrovascular bifurcations using geometrical features extracted from bifurcations. The method consists of five steps: image preprocessing, geometrical characteristics extraction, dimensionality reduction, machine learning model construction, and evaluation of classification models. The various phases of the proposed methodology are depicted in Fig.~\ref{fig:schema_DR}.

\subsubsection{Preprocessing and feature computation} 

\begin{figure}[ht!] 
    \centering
  \subfloat[ \label{BET}]{%
       \includegraphics[width=0.4\linewidth,height=2.7cm]{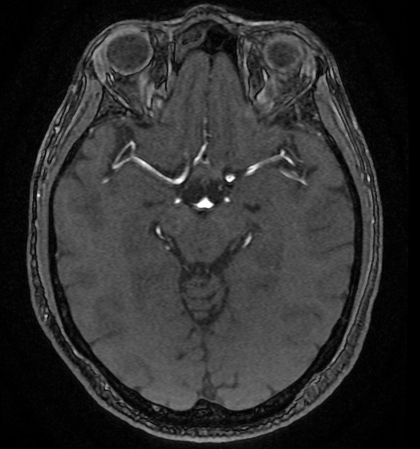}}
    \hfill
  \subfloat[ \label{seg}]{%
        \includegraphics[width=0.4\linewidth,height=2.7cm]{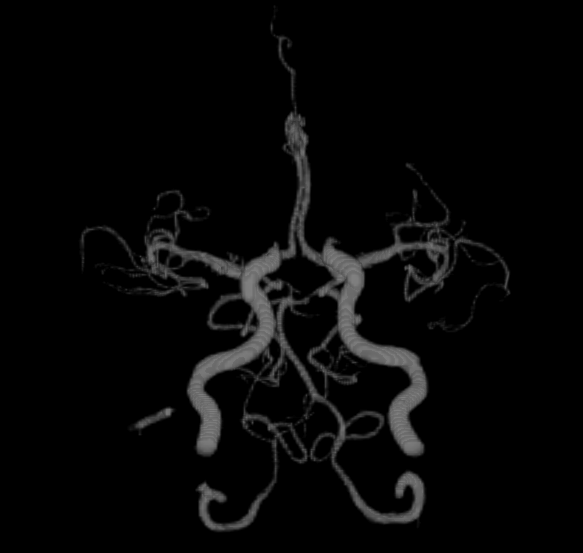}}
  \caption{\label{fig:Vesel_seg} (a) an MRA-TOF with (b), its segmentation. }
\end{figure}

Unlike previous works, mostly focusing on the classification of already selected bifurcations; in this work, we consider the entire MRA-TOF, and hence, no preselection is conducted on the bifurcations. To accurately extract the geometric features of the vascular tree, the arteries in the 3D MRA  (Fig.~\ref{fig:Vesel_seg}(a)) images first go through skull stripping, and are then segmented (Fig.~\ref{fig:Vesel_seg}(b), then the bifurcations coordinates within the brain vasculature are collected \cite{b15}.
Afterwards, various geometrical features are extracted onto the bifurcations and their branches~\cite{b9}: tortuosity, diameters, angles and lengths in addition to various combinations. Overall, 61 features are used to describe each bifurcation~\cite{b9}. 

\subsubsection{Dimensionality reduction} 

High-dimensional data presents significant challenges in a wide range of machine learning applications. Such techniques attempt to address these issues by transforming data into a lower-dimensional representation while preserving the critical information. Linear Discriminant Analysis (LDA), Principal Component Analysis (PCA), and Isomap are popular dimensionality reduction algorithms. While projecting the data into a lower-dimensional space, the LDA seeks for linear features combinations maximizing class separability. Contrarily, the PCA locates the largest variance among the data and projects them onto a subspace defined by the principle components. The Isomap is a non-linear dimensionality reduction technique that constructs a low-dimensional embedding based on geodesic distances between data points in a high-dimensional space. Here, those methods are compared to determine their performance in classifying features within a lower dimensional space. 


\subsection{Classification based on Autoencoders}

\begin{figure*} 
 \begin{center}
 \includegraphics[height=7cm, width=14cm]{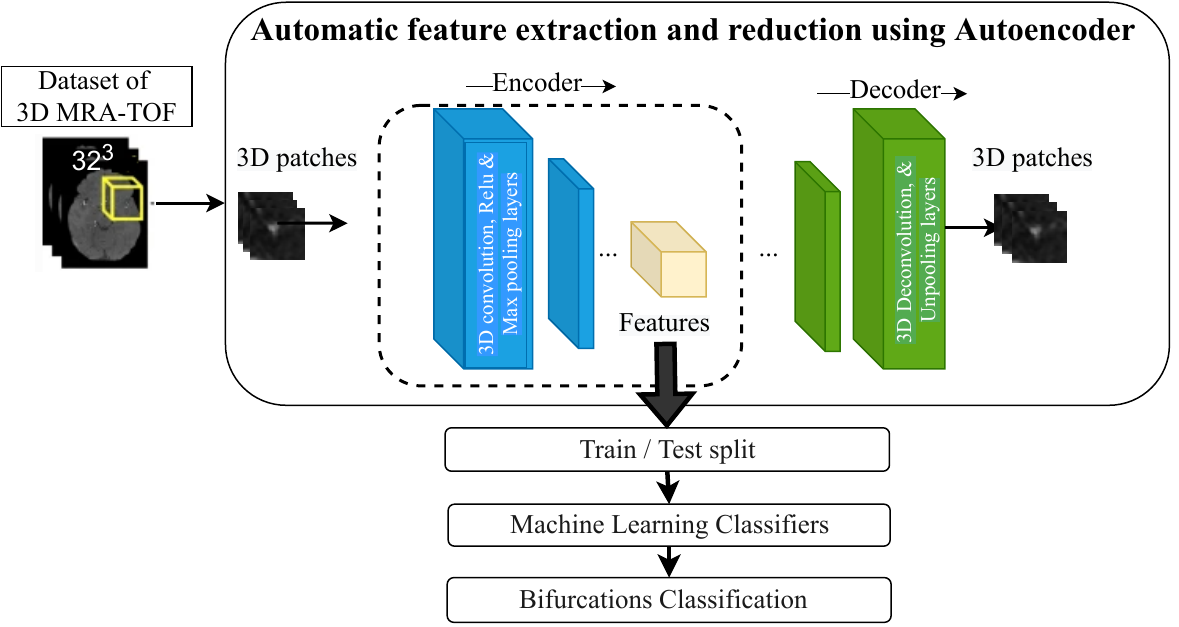}
	\caption{The proposed system model based on 3D CAE}
	\label{fig:schema_autoencoders}
 \end{center}
\end{figure*}

Automatic labeling of bifurcations in medical imaging can be modeled as a supervised learning algorithm. Specifically, the algorithm learns from labeled training bifurcations to make predictions and assigns labels to unseen data. The development of the classification algorithm consists of 4 main phases: \textit{i)} The preparation of the MRA-TOF data acquisitions and pre-processing steps, \textit{ii)} Feature extraction and dimensionality reduction using a 3D autoencoder, \textit{iii)} Use of the extracted features as input of different machine learning classifiers and \textit{iv)} Performance evaluation on a test bifurcation dataset using evaluation metrics. The general flowchart of the proposed classification algorithm is shown in Fig \ref{fig:schema_autoencoders}.

\subsubsection{MRA-TOF Preprocessing} 

After collecting the image dataset, several preprocessing steps are performed for each 3D image.  First, skull stripping is applied, followed by a vascular tree segmentation (see Fig. \ref{fig:Vesel_seg} (b)). Then, we use
the algorithm proposed by \cite{b15} in order to detect the bifurcations centers and extract their 3D coordinates. Next, based on the extracted 3D coordinates of the bifurcations, 3D gray level patches are collected within patches of size $32\times 32\times 32$ voxels. 

\subsubsection{Features extraction and dimensionality reduction} 
Along with recent advances in deep learning techniques based on convolutional neural networks, several studies have evaluated the applicability of a deep learning-based algorithm for cerebral vasculature analysis using MRA-TOF\cite{b10}. Convolutional AutoEncoders (CAE) are models based on neural networks, which have been developed for dimensionality reduction and representation learning in a variety of tasks\cite{b11}. The innovation of CAE for image analysis lies in the use of convolutional layers able to create an abstract representations of the original inputs by eliminating noise and redundant information\cite{b12}.


For 
features extraction, the encoder part of the CAE consists of convolution and pooling layers that progressively reduce the sampling of the input image. These layers learn to extract hierarchical features such as edges, textures, and top-level patterns from bifurcations crops. The compressed representation obtained by the encoder serves as a lower dimensional representation of the input image. This reduction in dimensionality captures the most relevant and discriminative information from the images while eliminating noise or irrelevant details. The size of the feature vector can be adjusted according to the desired level of compression.



In this experiment, a 3D autoencoder from \cite{b13} is used to learn features from the  bifurcations. The autoencoder is trained on a dataset of bifurcation patches, and the learned features are used as input to the classification step. We used 3D MRA-TOF patches as input to our network. It is important to note that the decoder of the CAE, which is responsible of the reconstruction step and the resampling of the compressed representation to the original size, was not used in our work. 


\section{Experimental results}
\subsection{Dataset}


The study included 91 patients selected from different scans and with different resolutions to compose our bifurcation dataset. We identified 13 bifurcations of interest (BoI) with an increased risk of developing an aneurysm, thus leading to a possible upper limit of $91\times13=1183$ bifurcations. 
However, due to a well known
\cite{b16} important variability among humans CoW anatomy, only $664$ BoIs were actually inventoried. Indeed it may happen that some arteries are either hypoplasic (underdevelopped and hence very thin) or aplastic (missing), and thus several bifurcations can be missing. 
%
In addition, a balanced number of classes is needed to accurately predict the correct class labels for new, unseen instances. For this reason, we have tried to approximate the number of bifurcations that constitute the CoW and the ones that are not of interest. Thus, the evaluation of our method was based on the $664$ BoIs and $664$ BNs randomly selected 
from the external zone around the CoW.

\subsection{Machine learning classification}

In the experiment based on the CAE, the encoder part of the network is only used for feature extraction and dimensionality reduction. 
The feature vector obtained from the encoder is used to feed a separate classification model. The classification model learns to associate the extracted features with the corresponding bifurcation labels. 
For the geometrical features-based method, LDA, PCA and Isomap were used to represent the bifurcation geometrical features in a lower dimensional space, the machine learning classifiers were then trained in the new components of this reduced space.

For the classification stage, various machine learning algorithms (MLAs), including Random Forest (RF), Naive Bayes (NB), Support Vector Machine (SVM), Quadratic Discriminant Analysis (QDA), Decision Tree (DT) and Multi-Layer Perceptron Neural Networks (MLP), were used to classify the bifurcations.

For unbalanced data sets, the stratified cross-validation (CV) evaluation method, produces more robust and reliable performance estimates. In our study, the approach involves dividing the dataset into 10 folds, while ensuring that each fold retains a proportional representation of the bifurcations classes. The model is trained on 9 folds containing a seminar number of bifurcations and tested on the remaining fold. This process is repeated 10 times, each fold being used once as a test. This ensures that the model is trained and tested on a balanced and representative sample of the data. Note that we have tested other evaluation methods, such as the Train-Test Split approach, although the CV performed well.

\subsection{Classification based on Autoencoders}

The proposed approach was evaluated on a dataset of 1328 intracranial bifurcation patches, of which 664 points constitute the CoW. As discussed before, 664 BNs were used to train the model finding True Negative bifurcations. 
We validate the effectiveness and robustness of the proposed method by performing full experimental simulations using various MLAs previously described . 

\begin{figure} [ht]
   \begin{center}
   \includegraphics[width=8cm, height=5cm]{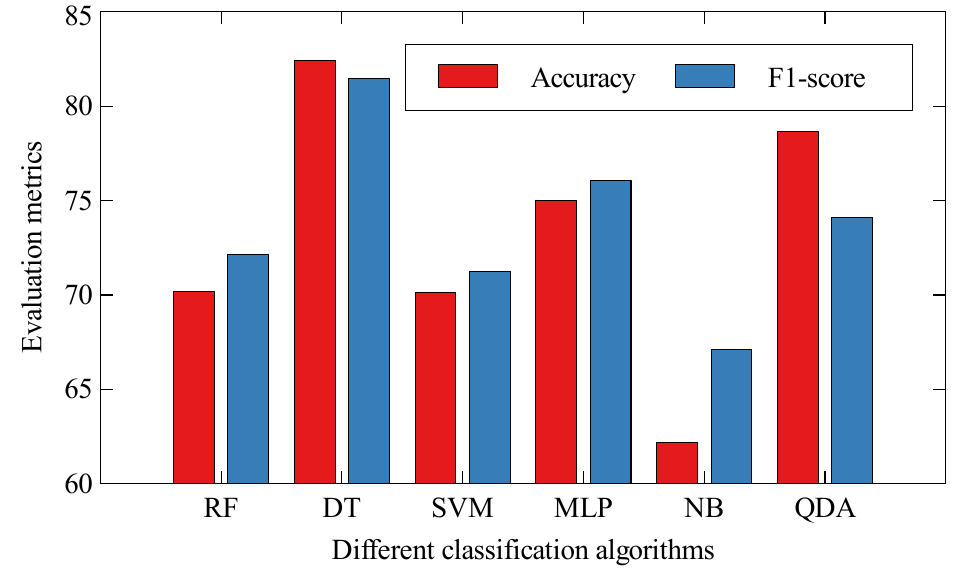}
   \end{center}
\caption{\label{fig:Accuracy_F1_method2}Comparison of different MLAs based on the Accuracy and F1 score (CAE based-method)}
\end{figure}

Fig.~\ref{fig:Accuracy_F1_method2} shows the results of average Accuracy and F1-score obtained on implementing the ten-CV. This figure illustrates that the highest classification was reached with the DT classifier, which provides an average accuracy and F1 score of 82.4\% and 81.47\% respectively. While the NB classifier gives a significant misclassification rate with an accuracy and F1-score around 60\% to 65\%.

%
According to the results obtained with the CAE-based method, the expected results were not achieved. One of the main reasons may be the unbalanced number of classes (especially BNs), which can lead to a bias in the extracted features. Another reason is that autoencoders may not be able to capture the full complexity of patches, leading to irrelevant or non-specific features that may not be useful for classification. 
Furthermore, this experiment does not take into consideration the 3D coordinates of the bifurcations, and hence, making the differentiation difficult between the BoIs and the BNs.

\subsection{Classification based on geometrical features}

\begin{figure}[ht!] 
    \centering
  \subfloat[Accuracy\label{AccuracyMice}]{%
       \includegraphics[width=0.5\columnwidth]{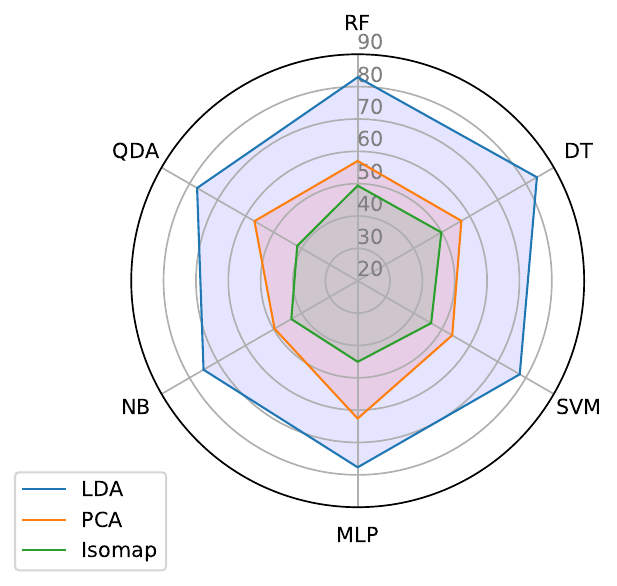}}
    \hfill
  \subfloat[F1-Score\label{F1ScoreMice}]{%
        \includegraphics[width=0.5\columnwidth]{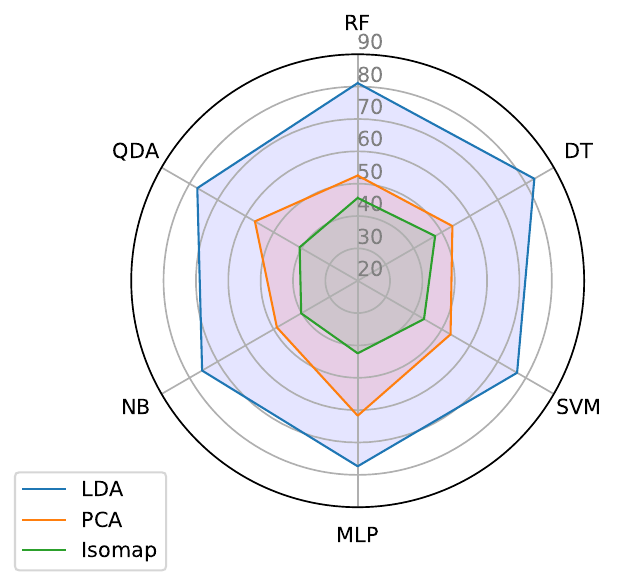}}
  \caption{\label{fig:RadarPlot}Performance of bifurcations identification
using multiple MLAs in terms of Accuracy and F1-score.}
\end{figure}

Similarly to the previous experiment, the 10-folds CV  method is used for a successful performance evaluation and to ensure a proper generalization of the MLAs. The Fig. \ref{fig:RadarPlot} presents the MLAs classification results achieved when applying different dimensionality reduction algorithms using the CV method. 

According to Fig. \ref{fig:RadarPlot}, the LDA is the  method achieving the highest classification rate while using the DT classifier. The average accuracy  and F1-score of LDA are 83.96\% and 83.04\% respectively. These results were obtained with 8 reduced components. Furthermore, the features provided by the PCA (10 compressed features) resulted in an average classification with all classifiers. However, MLAs applied on the Isomap reduced features(with 6 compressed features) showed significantly reduced performances.

\begin{figure} [ht]
   \begin{center}
   \includegraphics[width=8.3cm,height=6.5cm]{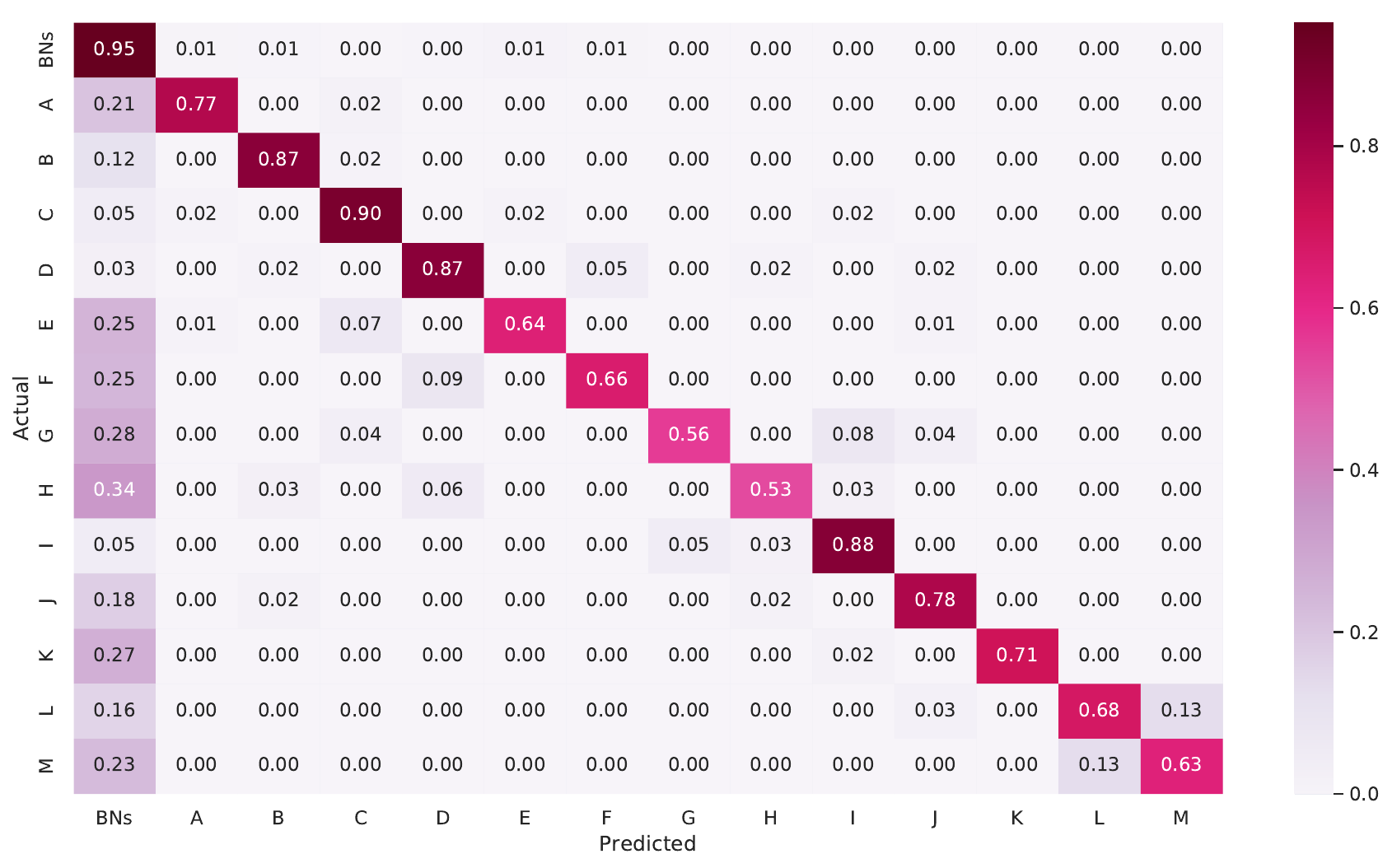}
   
   \caption{\label{fig:CM}Confusion matrix for 14-class classification implementing the DT classifier and LDA for DR. }
   \end{center}
\end{figure}

Furthermore, the confusion matrix is a very useful method for understanding the types of errors or correct predictions made by the model. 
Figure \ref{fig:CM} illustrates the confusion matrix for classification results retrieved with the DT algorithm on LDA reduced data. 
The imbalance in the training set means that DT 
is biased in favor of the majority groups. 
Indeed, we find that the majority of BoIs misclassifications are confused with BNs, as illustrated by the first column. This may be due to the complexity of the used database, with the BN class being more prevalent than the BoI category. In addition, some BN bifurcations are very close to BoIs, which leads to classification errors.

\section{Discussion and conclusion}In this work, we propose a method allowing the automatic labeling of the bifurcations of interest within TOF-MRA acquisitions. 
This paper presents a comparative analysis between classification based on vascularization geometry and connectivity features and classification based on extracted and reduced features of bifurcation patches using a convolutional autoencoder.

The first experiment illustrates how 3D autoencoders and machine learning classifiers can be jointly used to recognize the main bifurcations of the CoW from a full cerebral vasculature. The results suggest that the proposed methodology can be used for bifurcation classification issues, which can speed up the diagnosis and treatment of brain diseases via an automatic feature extraction. The proposed approach achieves an average classification accuracy above 82.4\% using a DT classifier. On the other hand, the second experiment based on the geometrical characteristics have been used to identify the bifurcations. The most accurate classification was achieved using the LDA + DT classifier solution showing an accuracy above 83.96\% when using the tenfold stratified CV evaluation method.

One explanation for algorithms limitations to produce improved results in terms of accuracy and F1 score can be explained as in some cases the bifurcations are very close together because the cerebral arteries are small, and therefore identification between the appropriate bifurcations is not easy and requires more knowledge. In addition, the difference between the number of BNs points and BoIs bifurcations contributes to misclassification of bifurcations obtained using MLAs. The class balancing can be achieved by various tasks, such as oversampling the minority class by generating more interest bifurcations forming the CoW, or reducing the majority classes (BNs). 

In summary, while autoencoders proved to be a promising approach in other areas, the results suggest that for cerebral bifurcation classification, exploiting geometric features leads to higher performance and more reliable results. Further research is needed to explore ways of combining the strengths of geometric features and deep learning techniques to potentially improve classification accuracy in bifurcation analysis. Also, future work will focus on expanding the dataset, exploring differents machine learning models, using external features  such as 3D coordinates of bifurcation center and its extremities to increase the identification rate using autoncoders. Thanks to their ability to learn efficient representations of the data, autoencoders can be trained on raw image data without explicitly segmenting the images or extracting 3D coordinates. Consequently, the use of autoencoders for bifurcation identification without a segmentation phase also offers an interesting prospect and potential for future research.

\end{document}